\documentclass{PoS}

\def\DU  {\mathop{{\cal D}\hbox{U}}}
\def\Dpsi {\mathop{{\cal D}\bar{\psi}{\cal D}\psi}}
\def\dd  {\mbox{d}}
\newcommand\detn[1]{\mbox{det}_{#1}}
\newcommand\tdetn[1]{\widetilde{\mbox{det}}_{#1}}
\def\Re {\mathop{\hbox{Re}}}
\title{Multi-histogram Reweighting for Finite Density Lattice QCD}
\ShortTitle{Reweighting method in finite density lattice QCD}

\author{\speaker{Anyi Li}, Andrei Alexandru, Keh-Fei Liu\\
        Department of Physics and Astronomy, University of Kentucky, Lexington KY 40506, USA\\
        E-mail: \email{anyili@pa.uky.edu},
        \email{alexan@pa.uky.edu},
        \email{liu@pa.uky.edu}}
\abstract{Finite density simulations require dynamical fermions
which are computationally demanding. We employ Ferrenberg and
Swendsen reweighting method to reduce the number of ensembles
needed. We use their method to do a combined reweighting in both
$\beta$ and quark number $k$ on an ensemble generated by the Kentucky group.}

\FullConference{XXIVth International Symposium on Lattice Field Theory\\
                July 23-28, 2006\\
                Tucson, Arizona, USA}
\begin{document}
\section{Introduction}
Finite density simulations require dynamical fermions which are
computationally demanding. When studying phase transitions we have
to scan very finely both the temperature and density directions; a
large number of ensembles need to be generated. Ferrenberg and
Swendsen~\cite{fs88}~\cite{fs89} showed that we can employ
reweighting to reduce the number of ensembles needed. Using their
methods we perform an extrapolation in $\beta$ and then we do a
combined reweighting in both $\beta$ and the quark number $k$. We
take the ensembles generated at $k=0$ and $k=3$, and used them to
extrapolate to $k=6$ case. The results of reweighting are also
compared against the ones of that direct measurements.

\section{Simulation Background}
Finite density ensembles were generated using the canonical
approach~\cite{afhl05}. To build canonical partition function,we start from
fugacity expansion of the grand canonical partition function,
\begin{equation} Z(V,T,\mu) = \sum_{k} Z_C(V, T, k) e^{\mu k/T},
\end{equation}
where $k$ is the net quark number, $Z_C$ is the canonical
partition function of the system. On a lattice, we can easily
compute the Fourier transform of the grand canonical partition
function
\begin{equation}
Z(V,T,\mu) = \int \DU \Dpsi e^{-S_g(U)- S_f(\mu; U,
\bar{\psi},\psi)}
\end{equation}
to get the canonical partition function
\begin{equation}
Z_C(V, T, k) = \frac{1}{2\pi} \int_0^{2\pi} \mbox{d}\phi \,e^{-i k
\phi} Z(V, T, \mu)|_{\mu=i\phi T}.
\end{equation}
We will specialize to the case of two degenerate flavors. After
integrating out the fermion part, we get a simple expression
\begin{equation}
Z_C(V, T, k) = \int \DU e^{-S_g(U)} \detn{k}
M^2(U)\label{eq:canonical},
\end{equation}
where
\begin{equation}
\detn{k} M^2(U) \equiv \frac{1}{2\pi}\int_0^{2\pi} \dd\phi\,e^{-i
k \phi} \det M(m, \mu;U)^2|_{\mu=i\phi T}
\end{equation}
is the determinant projection to the fixed net quark number $k$.
Then $Z_C$ is used to generate the ensembles at particular $k$.

Simulations based on the action give in~(\ref{eq:canonical})
cannot be carried out directly since the integrand is not positive
definite. The determinant is split in two parts: a positive
definite part that is used to generate the ensemble and a phase.
For a more detailed presentation we refer the reader to the
original paper~\cite{afhl05}.

The absolute value of the Polyakov loop is given by
\begin{equation}
\left< |P| \right> = \frac{ \left< |P| \alpha \right>_0} {\left<
\alpha \right>_0},
\end{equation}
where
\begin{equation}
\alpha(U) = \frac{\tdetn{k} M^2(U)}{\left|\Re \tdetn{k}
M^2(U)\right|},~~\tdetn{k} M^2(U) \equiv \frac{1}{N} \sum_{j=0}^{N-1}
e^{-i k \phi_j} \det M(U_{\phi_j})^2 
\end{equation}
is the phase and $\left<\right>_o$ stands for the average over the
ensemble generated with measure $\left|\Re \tdetn{k}
M^2(U)\right|$. Quark chemical potential is defined by the
expression below:
\begin{eqnarray}
\mu(k) &=&
-\frac{1}{\beta}\ln\frac{\tilde{Z}_C(k+1)}{\tilde{Z}_C(k)} =
-\frac{1}{\beta}  \ln \frac{1}{\tilde{Z}_C(k)} \int \DU e^{-S_g(U)} \nonumber \\
&\times&\frac{1}{N}\sum_{j=0}^{N-1} e^{-i\phi_j} e^{-i k \phi_j}
\det M^2(U_{\phi_j}) \nonumber \\
&=& -\frac{1}{\beta}\ln \left<e^{-i\phi}\right>_k.
\end{eqnarray}
More relevant to our study is the baryon chemical potential:
\begin{equation}
\mu_B(n_B) = -\frac{1}{\beta} \ln\left< e^{-i3\phi}\right>_{3
n_B}.
\end{equation}

 To test the reweighting method, a subset of the original
ensembles was employed and the results of the extrapolation
were compared against the results of the direct measurements.

\section{Algorithm}
  The original reweighting method~\cite{fs88} employed an ensemble generated at a particular point
in the parameter space to build an ``induced'' ensemble for a
different value of parameters. The method is limited to a
neighborhood of the original point in the parameter space. At
distant points the extrapolation becomes unreliable due to poor
overlap between the original and the target distribution.

  The multi-histogram method~\cite{fs89} addresses the limitations of the original method by employing several ensembles.
A carefully chosen set of ensembles, covering the area of interest
in the parameter space, allows us to interpolate reliably. Thus,
we can finely scan the area of interest with only a small set of
ensembles.

  The basic idea behind the reweighting method is to use the histogram generated by the simulated ensembles as an approximation of the
probability distribution; this approximation is then used to
approximate the probability distribution at a different point in
the parameter space. Once we have an approximation for the
probability distribution we can compute all observables.

We used these methods in two scenarios:
\begin{enumerate}
\item Fix $k$, vary $\beta$.
 \begin{itemize}
  \item Single histogram method: The histogram of only one ensemble is used in this case.
  \item Multi-histogram method: The histograms of several ensembles generated at different points are combined.
 \end{itemize}
\item Vary both $\beta$ and $k$. (Only Multi-histogram method is used.)

\end{enumerate}

  In our first case (fix $k$, vary $\beta$), we approximate the
  ``weight'' of each configuration $U$ by
\begin{equation}
P(U,\beta) =
\frac{e^{-S_{\beta}(U)}}{\sum_{m=1}^{R}n_{m}e^{-S_{\beta_m}(U)-f_m}}\label{eq:a}
\end{equation}
This approximation allows us to compute averages at arbitrary
values of $\beta$. The input for this formula consists of $R$
ensembles generated by weights given by $S_{\beta_1}, \ldots,
S_{\beta_R}$ with $n_1, \ldots, n_R$ configurations in each
ensemble.

To use the formula above we need to determine the parameters
$f_n$~---~they are the ``free energies'' of each ensemble $n$,
\begin{equation}
exp(f_n) = \sum_{U_i}P(U_i,\beta_n)\label{eq:b}
\end{equation}
where we sum over all the configurations $U_i$ in the ensemble.
Eq.~(\ref{eq:a}) and ~(\ref{eq:b}) are used to determine the free
energy $f_n$. We start with a guess for $f_n$ and plug this into
Eq.~(\ref{eq:a}) to get an approximation for the probability
distribution $P(U,\beta)$. Then, we take this approximation for
$P(U,\beta)$ and plug it into Eq.~(\ref{eq:b}) to get a more
refined guess of $f_n$. We repeat this process until $f_n$
converges.

After convergence is reached, we can measure the observables at
different $\beta$ by
\begin{equation}
 <O>_{\beta}=\frac{\sum_{U}O(U)P(U,\beta)}{\sum_{U}P(U,\beta)}
\end{equation}

  For our second case (both $\beta$ and $k$ are varied), the ``weight'' for different configurations at different $k$ and $\beta$ should be
calculated as
\begin{equation}
P(U,k',\beta ') = \frac{|\mathop{\hbox{Re}} \mbox{det}_{k'}M^2(U)|
e^{-S_{\beta '}(U)}}
{\sum_{i=1}^{R}n_{i}|\mathop{\hbox{\hbox{Re}}} \mbox{det}_{k_i}
M^2(U)|e^{-S_{\beta_i}(U)-f_i}}
\end{equation}
\begin{equation}
\exp(-f_i) = \sum_{U_j} P(U_j,k_i,\beta_i)\label{eq:f}
~~~\mbox{\small We sum over all configurations}
\end{equation}

The chemical potential can be measured by using the ``weight'' $P(U,k,\beta)$:
\begin{eqnarray}
 \beta \mu_B(n_B,\beta) & = & - \mbox{ln}
                              \frac{1}{Z_c(3n_B,\beta)} \int {\cal D} U e^{-S_{\beta}(U)} \nonumber \\
                          & & \times \frac{1}{N} \sum_{j=0}^{N-1} e^{-i3\phi_j}e^{-i3n_B\phi_j}
                                \mbox{det}M^2(U_{\phi_j})
\end{eqnarray}

\subsection*{Optimized convergence}

  In our calculations, we found that the iteration presented above converges very slowly.
For a more efficient convergence we use the following procedure:
we know that $\vec{f}^{(k+1)}$ (the $k+1^{th}$ iteration step) can
be expressed as a function of $\vec{f}^{(k)}$ ($k^{th}$ step)
\begin{equation}
\vec{f}^{(k+1)} = \vec{G}(\vec{f}^{(k)}) ~~~\vec{G}~\mbox{is the
vector function of}~\vec{f}^{(k)}
\end{equation}
In our case, $\vec{f}$ is the free energy, $\vec{G}$ is the
recurrent equation~(\ref{eq:f}) for the free energy calculation.
The fixed point $\vec{f}^{(\infty)}$ has the property that
$\vec{f}^{(\infty)}=\vec{G}(\vec{f}^{(\infty)})$. At step $k$ our
approximation is $\vec{f}^{(k)}$, we write then:
\begin{equation}
 \vec{f}^{(k)} + \delta{\vec{f}^{(k)}} = \vec{G}(\vec{f}^{(k)} +
 \delta{\vec{f}^{(k)}})~~~\mbox{where}~~~\delta{\vec{f}^{(k)}}=\vec{f}^{(\infty)}-\vec{f}^{(k)}.
\end{equation}
Assuming that $\vec{f}^{(k)}$ is close to $\vec{f}^{(\infty)}$
then $\delta{\vec{f}^{(k)}}$ is small and we can approximate
\begin{equation}
\vec{G}(\vec{f}^{(k)} +
 \delta{\vec{f}^{(k)}})\approx\vec{G}(\vec{f}^{(k)})+\triangledown\vec{G}\cdot\delta{\vec{f}^{(k)}}.
\end{equation}
Using this we get:

\begin{equation}
f_i^{(k)} + \delta{f_i^{(k)}} \approx G_i(\vec{f}^{(k)}) +
\frac{\partial{G_i}}{\partial{f_j^{(k)}}}\delta{f_j^{(k)}}
\end{equation}

$\frac{\partial{G_i}}{\partial{f_j^{(k)}}}$ is the element of
matrix $M^{(k)}=\frac{\partial{G_i}}{\partial{f_j^{(k)}}}$,
\begin{equation}
f_i^{(k)} + \delta{f_i^{(k)}} \approx G_i(\vec{f}^{(k)}) +
M_{ij}^{(k)}\delta{f_j^{(k)}}
\end{equation}
so
\begin{equation}
(I - M^{(k)})_{ij}\delta{f_j^{(k)}} \approx G_i(\vec{f}^{(k)}) -
f_i^{(k)}
\end{equation}
\begin{equation}
\delta{f_j^{(k)}} \approx (G_i(\vec{f}^{(k)})-f_i^{(k)}) \cdot {(I
- M^{(k)})^{-1}}_{ij}
\end{equation}

We expect the matrix $(I - M^{(k)})$ to be almost singular when we
are  approaching fixed point, so we use SVD to compute the matrix
inversion. Our next step in the iteration is then
$\vec{f}^{(k+1)}=\vec{f}^{(k)}+\delta\vec{f}^{(k)}$ rather than
$\vec{G}(\vec{f}^{(k)})$. Using this method the convergence of the
iteration is greatly accelerated.
\section{Results and Conclusion}
We determine the Polyakov loop using the single histogram method and the multi-histogram method and compare the results with the direct
measurements. The results are plotted as a function of $\beta$, the gauge coupling constant.

\begin{figure}[htbp]
\begin{center}
\includegraphics[width=0.43\textwidth]{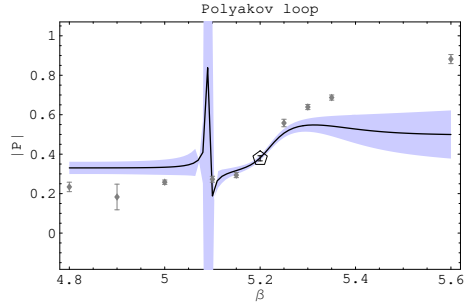}
\end{center}
\caption{Polyakov loop (Single histogram at fixed $k=3$). All dots
represent the ensembles where direct measurements were carried
out. The ones surrounded with polygon are used as our reweighting
input; the gray ones are not. The line represents the result of
the extrapolation and the band represents the statistical errors.}
\end{figure}

From single histogram reweighting of Polyakov loop, we can see from Fig.~1 
the method is limited to a neighborhood of the original point
$\beta=5.20$ in the parameter space. The extrapolation at distant
points becomes unreliable due to poor overlap. The multi-histogram
method addresses this limitation, and by employing several
ensembles, we expect to interpolate results reliably.
 \clearpage
\begin{figure}[ht]
\begin{center}
\begin{tabular}{cc}
  a & b \\
  \includegraphics[width=0.42\textwidth]{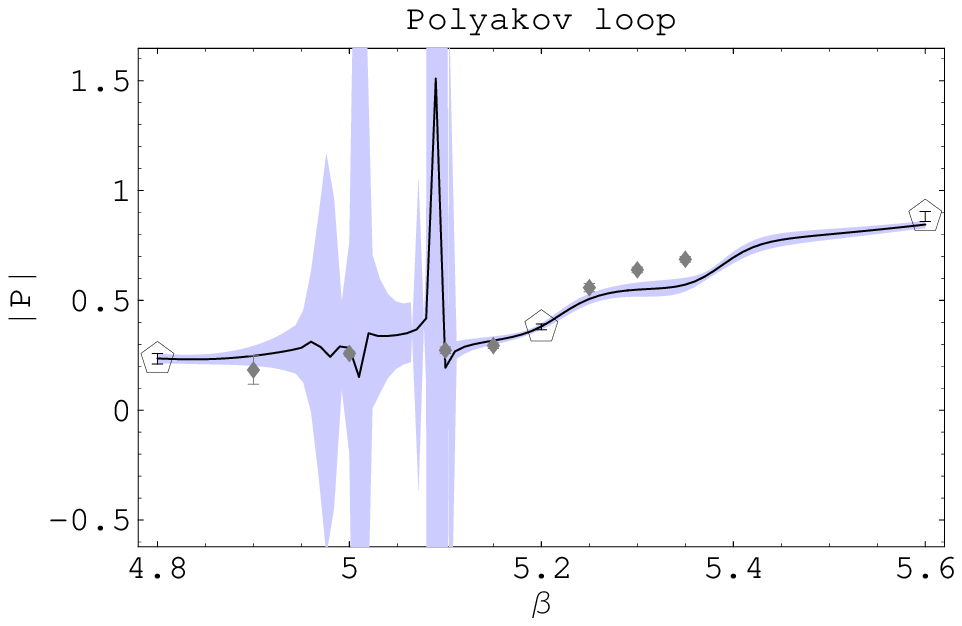} & \includegraphics[width=0.42\textwidth]{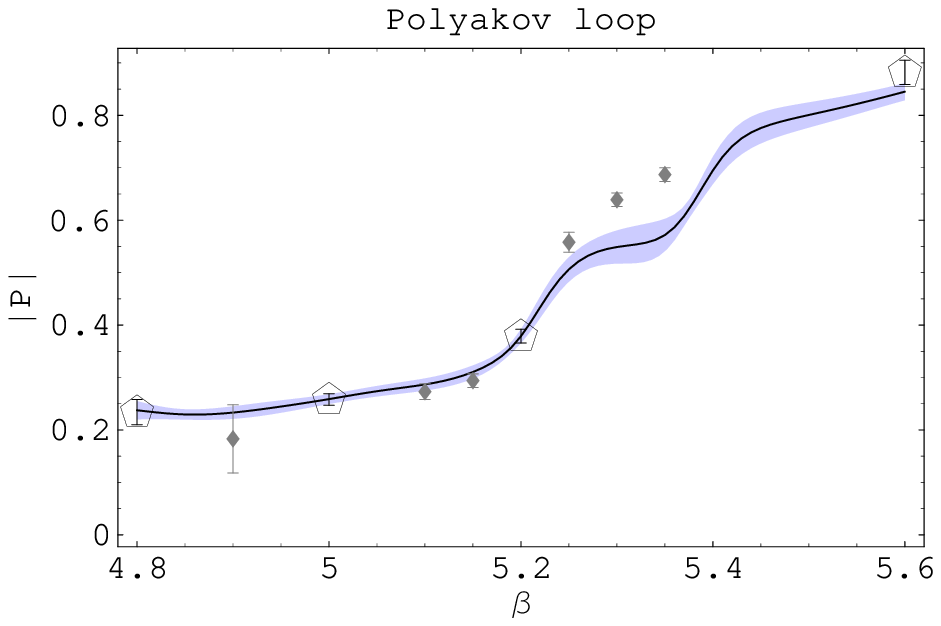} \\
  c & d \\
  \includegraphics[width=0.42\textwidth]{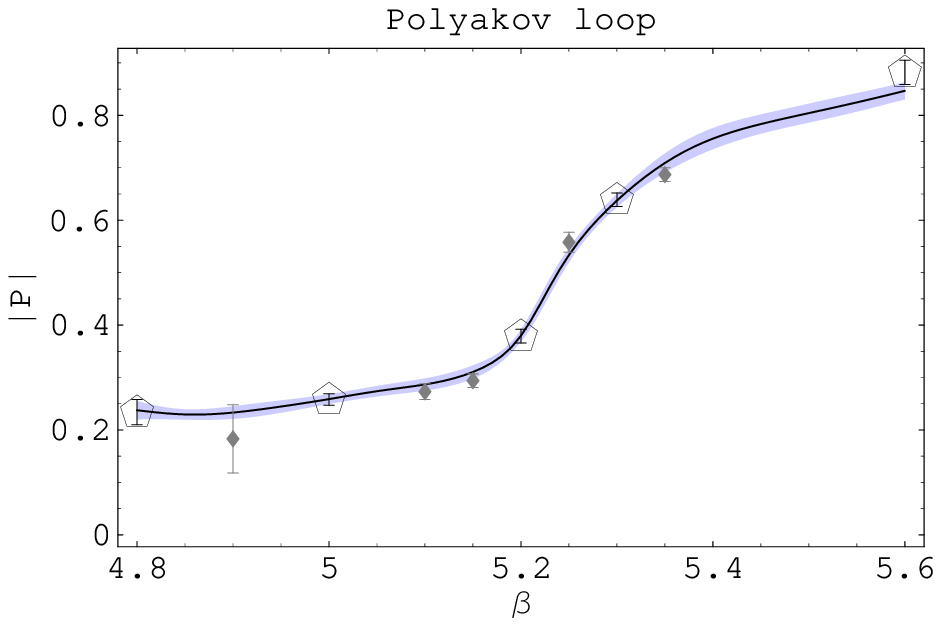} & \includegraphics[width=0.42\textwidth]{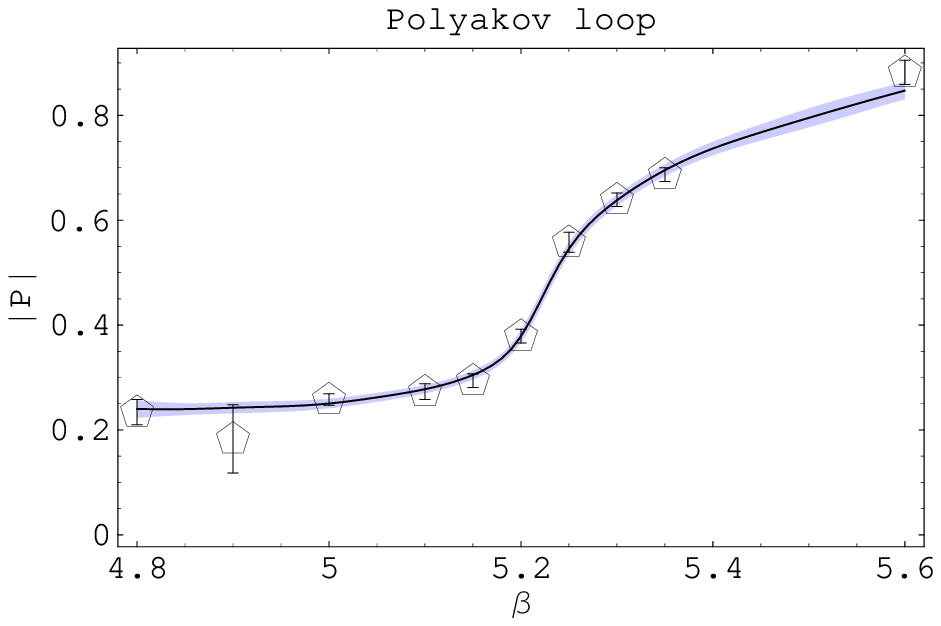} \\
\end{tabular}
\end{center}
\caption{Polyakov loop(Multi-histogram, fixed $k=3$). Same as in Fig.~1 for multi-histogram reweighting with k=3. From a to b,
more ensembles are added.}
\end{figure}

  We start by employing three ensembles in Fig.~2. Two of them at $\beta=4.8$ and
  $\beta=5.6$ are the bounds of the region of interest and the third
  one is in the middle at $\beta=5.2$. We plot the results in
  Fig.~2a; we see that the error bars are quite large between
  $\beta=4.9$ and $5.1$. When we add one more ensemble in the
  ``problem'' region, the plot changes significantly as we can see
  from Fig.~2b. The largest error bars are around $\beta=5.3-5.4$
  now. Once we add a point at $\beta=5.3$ the error bars are reduced
  as we see from Fig.~2c. From this plot we also see that with only
  half of the generated ensembles the reweighting curve describes very well
  the data in the region of interest. From Fig.~2d, we see that adding
  more ensembles~(total 10 ensembles) doesn't change the curve significantly~---~the only difference is
  that the error band gets smaller.
\begin{figure}[b]
\begin{center}
\begin{tabular}{cc}
 \includegraphics[width=0.43\textwidth]{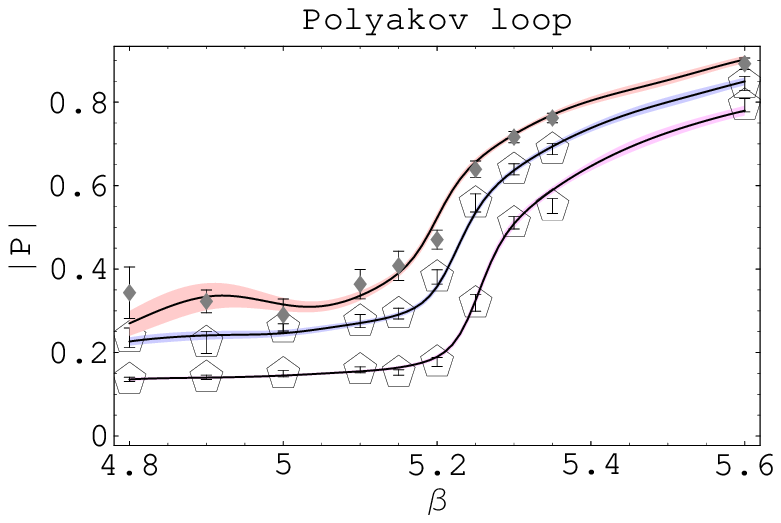} & \includegraphics[width=0.44\textwidth]{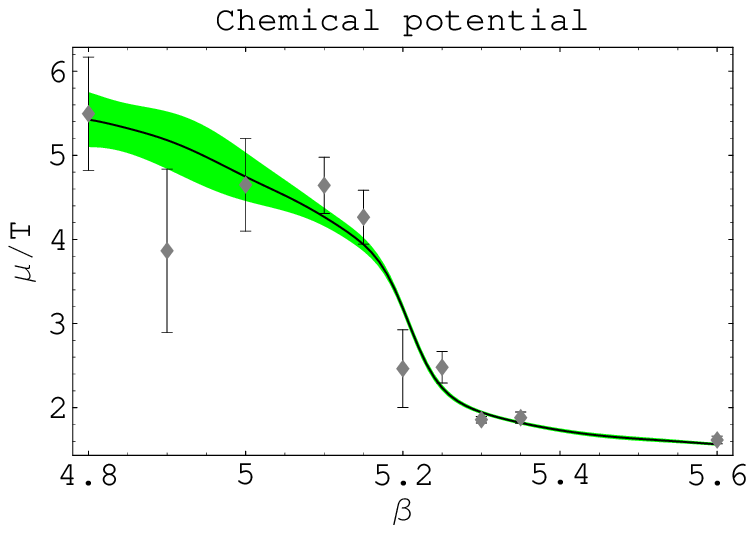}
\end{tabular}
\end{center}
\caption{We used ensembles generated at $k=0$ and $k=3$ to
extrapolate $k=6$ case, and the plot is compared to the direct
measurements.}
\end{figure}

We have also performed a combined reweighting in $\beta$ and $k$ in Fig.~3.
We employed ten ensembles at $k=0$ and ten ensembles at $k=3$ to extrapolate
to $k=6$. We used this method to compute both the Polyakov loop and the chemical
potential. As we can see from Fig.~3 the extrapolated curve matches rather well the direct measurements.

 We find that multi-histogram method can be used
effectively to reduce the number of ensembles needed for our
simulations: for our $\beta$ interpolations we find that we could
use as little as half of the ensembles generated to get a much
better scan of the temperature range. The way we envision using
this method is to generate few ensembles of first scanning the
temperature range rather coarsely and then add some more ensembles
at the points where ``features'' like sharp transitions appear in
our plots. The process stops when adding new ensembles does not
alter our plots significantly. We also find that interpolations in
both temperature and quark number are feasible and we plan to
employ these methods in our further studies.

\section*{Acknowledgments} The work is partially supported by DOE
Grants No. DEFG05-84ER40154 and No. DE-FG02-95ER40907.

\end{document}